\documentclass[a4paper,11pt]{article}
\usepackage{pos}

\newcommand{\lr}[1]{ \left( #1 \right) }
\newcommand{\lrs}[1]{ \left[ #1 \right] }

\newcommand{\vev}[1]{ \langle \, #1 \, \rangle }

\newcommand{\ket}[1]{ \, | #1 \rangle }
\newcommand{\bra}[1]{ \langle #1 | \, }

\newcommand{\tr}{ {\rm Tr} \, }

\title{Quantum chaos in supersymmetric Yang-Mills-like model: equation of state, entanglement, and spectral form-factors}

\ShortTitle{Quantum chaos in supersymmetric Yang-Mills-like model}

\author*[a]{Pavel Buividovich}

\affiliation[a]{Department of Mathematical Sciences, University of Liverpool, L69 7ZL Liverpool, UK}

\emailAdd{pavel.buividovich@liverpool.ac.uk}

\abstract{We analyze in detail a sharp transition between the low-energy, low-dimensional eigenstates and the high-energy chaotic bulk of the spectrum for a simple supersymmetric quantum-mechanical model with Hamiltonian $\hat{H}_S = \left(\hat{p}_1^2 + \hat{p}_2^2 + \hat{x}_1^2 \, \hat{x}_2^2\right) \otimes I + \hat{x}_1 \otimes \sigma_1 + \hat{x}_2 \otimes \sigma_3$, which mimics the structure of the Banks-Fischler-Susskind-Stanford (BFSS) matrix model, the spatially compactified $\mathcal{N} = 1$ super-Yang-Mills theory. We conjecture that this transition might be similar to the transition between the $D0$-brane and $M$-theory regimes in the BFSS model, and find that it does not lead to irregularities in the thermodynamic equation of state. We demonstrate that real-time spectral form-factor for our supersymmetric model exhibits the ``ramp'' behavior typical for quantum chaos. We also analyze the entanglement entropy and the spectrum of the reduced density matrix of the eigenstates of $\hat{H}_S$, considering one of the bosonic degrees of freedom as a subsystem. The entanglement entropy of low-energy eigenstates appears to be practically energy-independent. Exactly at the onset of random-matrix-type level spacing fluctuations, this behavior rapidly changes into a steady growth of entanglement with energy. We demonstrate that the spectrum of the reduced density matrix also exhibits universal level-spacing fluctuations towards its higher end, even for the ground state of the supersymmetric model. Thus even the regularly spaced, non-chaotic eigenstates contain some information about semi-classical chaotic dynamics at high energies.}

\FullConference{%
  The 39th International Symposium on Lattice Field Theory (Lattice2022),\\
  8-13 August, 2022 \\
  Bonn, Germany
}

\graphicspath{{./plots/},{./plots_proceedings/}}

\begin{document}
\sloppy
\maketitle

\section{Introduction and overview}
\label{sec:intro}

Chaotic real-time dynamics of Yang-Mills theory has been in the focus of intensive research since 1980's \cite{Savvidy:83}. Classical chaos of Yang-Mills fields was discussed as one of the potential mechanisms for fast apparent thermalisation in heavy-ion collisions \cite{Kunihiro:1008.1156,Schafer:0809.4831}, with the conclusion that the classical dynamics alone is not fast enough to agree with realistic estimates of thermalization times.

There has been a renewed interest to chaos in Yang-Mills-like models driven by ideas from holographic AdS/CFT duality. AdS/CFT duality maps thermalization process in supersymmetric Yang-Mills theory to the formation of a black hole in AdS space, with the Hawking temperature of a black hole being identified with the temperature of Yang-Mills fields \cite{Rajagopal:1101.0618,Schaefer:1012.4753}. We can therefore consider supersymmetric Yang-Mills theory as a microscopic model of AdS black hole dynamics.

Once Yang-Mills fields have thermalized, or, in a holographic dual description, once a black hole is formed, we can also ask how fast the thermalized system absorbs infinitely small perturbations. An equivalent question is how fast a black hole could scramble small bits of information thrown into it (e.g. a book thrown into a black hole of solar mass). Various thought experiments suggest that black holes are the fastest possible scramblers of information \cite{Susskind:0808.2096}, which implies, via holographic duality, that supersymmetric Yang-Mills theory should also be maximally chaotic. At a more technical level, propagation or scrambling of small perturbations in quantum systems can be characterized in terms of the so-called Out-of-Time-Order Correlators (OTOCs) $C\lr{t} = - \vev{ \lrs{\hat{A}\lr{t}, \hat{B}\lr{0}}^2 }$, where the time-dependent operator $\hat{A}\lr{t}$ is some measurable quantity, and the operator $\hat{B}\lr{0}$ perturbs the ground state of the system at time $t = 0$. For chaotic systems, the OTOCs grow exponentially over some time range: $C\lr{t} \sim e^{2 \lambda_L \, t}$, where $\lambda_L$ is the Lyapunov exponent. Maldacena, Stanford and Shenker \cite{Maldacena:1503.01409} demonstrated that for a wide class of quantum systems with parametrically many degrees of freedom, the Lyapunov exponents obey the universal bound $\lambda_L \leq 2 \pi \, T$. Correspondingly, ``maximally chaotic systems'' are systems that saturate the bound. In the holographic dual description, the MSS bound is usually saturated for black hole geometries. However, so far the only system for which the saturation of the MSS bound could be explicitly demonstrated from the field theory side is the Sachdev-Ye-Kitaev (SYK) model of interacting Majorana fermions \cite{Stanford:1604.07818}.

Besides the well-known $\mathcal{N} = 4$ supersymmetric Yang-Mills theory, one of the limits of supersymmetric Yang-Mills theory with a particularly well understood holographic dual description is the Banks-Fischler-Shenker-Susskind (BFSS) model \cite{Susskind:hep-th/9610043}, spatially compactified $\mathcal{N} = 1$ supersymmetric Yang-Mills theory in $d = 1 + 9$ dimensions. Depending on the scaling limit of model parameters, it is holographically dual to black $D0$-branes in type IIA superstring theory, or a Schwarzschild black hole in $M$-theory \cite{Maldacena:hep-th/9802042,Hanada:2110.01312,Costa:1411.5541}. Due to its relative simplicity, the BFSS model is convenient for numerical studies of thermodynamics \cite{Hanada:2110.01312,Hanada:1802.02985,Rinaldi:1606.04951,Hanada:2110.01312} as well as real-time dynamics \cite{Hanada:1602.01473,Buividovich:18:3,Buividovich:22:2}.

Of particular interest is the transition between $D0$-branes in type IIA superstring theory and the Schwarzschild black hole in $M$-theory \cite{Rinaldi:1606.04951,Maldacena:hep-th/9802042,Hanada:2110.01312}. The temperature of this transition is expected to decrease down to zero as the dimension of $SU\lr{N}$ gauge group grows and approaches the large-$N$ limit \cite{Maldacena:hep-th/9802042}. Numerical studies suggest that the Schwarzschild black hole/$M$-theory regime is indeed absent at large $N$. On the other hand, a recent study \cite{Hanada:2110.01312} found signatures of the $M$-theory regime in metastable states of Monte-Carlo simulations at low temperatures and finite $N$. As any other black hole, the Schwarzschild black hole in $M$-theory can be expected to be maximally chaotic and saturate the MSS bound on Lyapunov exponent. However, it is not clear what could be the mechanism of this saturation. To be consistent with the MSS bound, the Lyapunov exponent $\lambda_L$ should scale with the temperature $T$ as $T^\alpha$ with $\alpha \geq 1$. On the other hand, a straightforward scaling analysis suggests that classical chaotic dynamics leads to the scaling $\lambda_L \sim T^{1/4}$ \cite{Hanada:1512.00019,Buividovich:18:3,Buividovich:22:2}. Therefore, some mechanism other than classical chaos should be responsible for the saturation of the MSS bound at low temperatures. At low temperatures, the BFSS model is very strongly coupled. In the absence of exact analytic solutions, we can only use numerical analysis to obtain some results in this regime. While this is certainly possible for thermodynamic quantities \cite{Hanada:2110.01312,Hanada:1802.02985,Rinaldi:1606.04951,Hanada:2110.01312}, real-time dynamics of the BFSS model at sufficiently large $N$ is currently inaccessible to any numerical method.

We can only hope to be able to get some non-perturbative results on real-time behavior of the BFSS-like models for the smallest possible numbers of colors $N = 2$ or $N=3$ and the smallest number $d = 2$ of spatial dimensions. With the current state of technology, the most obvious practical method for simulating the real-time dynamics of the BFSS model with $d = 2$ and $N = 2$ is the numerical exact diagonalization \cite{Rinaldi:2108.02942,Buividovich:22:2}. Simulations on quantum computers might be possible as well, but are beyond the capabilities of modern quantum devices \cite{Rinaldi:2108.02942}. Even for $d = 2$, $N = 2$, the Hamiltonian of the BFSS model involves 6 bosonic and 3 fermionic degrees of freedom. Truncating the bosonic degrees of freedom to $\Lambda$ harmonic oscillator states, we are led to the Hilbert space with overall size $8 \Lambda^6$, which becomes prohibitively expensive even for $\Lambda \gtrsim 5$ \cite{Rinaldi:2108.02942}.

In the early days of the BFSS model/$M$-theory, de Wit, Lüscher and Nicolai \cite{Nicolai:NPB1989} introduced a significantly simpler supersymmetric Hamiltonian that captures the most important features of the full BFSS model \footnote{In the original work \cite{Nicolai:NPB1989}, the supersymmetric Hamiltonian was written as $\hat{H}_S = \hat{H}_B \otimes I + \hat{x}_1 \otimes \sigma_1 + \hat{x}_2 \otimes \sigma_2$. In these Proceedings, we follow \cite{Buividovich:22:2} and use a global unitary transformation to represent $\hat{H}_S$ in a manifestly real form (\ref{HS}).}:
\begin{eqnarray}
\label{HS}
 \hat{H}_S = \hat{H}_B \otimes I + \hat{x}_1 \otimes \sigma_1 + \hat{x}_2 \otimes \sigma_3
 =
 \left(
   \begin{array}{cc}
     \hat{H}_B + \hat{x}_2 & \hat{x}_1 \\
     \hat{x}_1             & \hat{H}_B - \hat{x}_2 \\
   \end{array}
 \right) ,
 \\
\label{HB}
 \hat{H}_B = \hat{p}_1^2 + \hat{p}_2^2 + \hat{x}_1^2 \, \hat{x}_2^2 .
\end{eqnarray}
This Hamiltonian contains only two bosonic degrees of freedom $\hat{x}_1$ and $\hat{x}_2$ (with the canonical conjugate momenta $\hat{p}_1$ and $\hat{p}_2$), and a single fermionic degree of freedom with a two-dimensional Hilbert space. The Pauli matrices $\sigma_1$ and $\sigma_3$, or, equivalently, the $2 \times 2$ block matrix in the last equality act on this two-dimensional Hilbert space. The potential energy term of the form $\hat{x}_1^2 \, \hat{x}_2^2$ mimics the non-Abelian commutator term $\tr\lr{\lrs{A_{\mu},A_{\nu}}^2}$ in the Yang-Mills action, and the terms $\hat{x}_1 \otimes \sigma_1 + \hat{x}_2 \otimes \sigma_3$ mimic the coupling of gauge fields to fermions. With considerably less degrees of freedom than for the $SU\lr{2}$ BFSS model, the model (\ref{HS}) can be studied numerically with significantly higher precision.

The supersymmetric Hamiltonian (\ref{HS}) is the simplest supersymmetric extension of the bosonic Hamiltonian $\hat{H}_B = \hat{p}_1^2 + \hat{p}_2^2 + \hat{x}_1^2 \, \hat{x}_2^2$, which was extensively studied in the literature as one of the simplest Hamiltonian systems that feature chaotic classical dynamics at all energies \cite{Savvidy:NPB84,Arefeva:hep-th/9710032}. It should be mentioned that the bosonic model is strictly speaking not ergodic and admits some stable periodic orbits \cite{DahlqvistPhysRevLett1990,ZakrzewskiPhysLettB1994} and non-universal distributions of energy level spacings for some of the irreducible representations of the $C_{4v}$ symmetry group \cite{Zakrzewski:chao-dyn/9501016}. The bosonic Hamiltonian $\hat{H}_B$ can be obtained as a zero angular momentum projection of $SU\lr{2}$ bosonic matrix model \cite{Sekino:1403.1392,Berenstein:1608.08972,Kares:hep-th/0401179}, which is a dimensional reduction of $SU\lr{2}$ pure Yang-Mills theory in $d = 1 + 2$ dimensions. A higher-dimensional generalization of this Hamiltonian was considered recently in \cite{Kolganov:2205.05663} using $1/N$ expansion techniques. The effects of Chern-Simons, BMN and ABJM deformations on classical chaos in dimensionally reduced Yang-Mills-like models were studied in detail in \cite{Baskan:2101.05649, Baskan:2203.08240,Baskan:1912.00932}.

After the original work of de Wit, Lüscher and Nicolai \cite{Nicolai:NPB1989}, the supersymmetric Hamiltonian (\ref{HS}) was studied numerically in \cite{Korcyl:hep-th/0610105,Wosiek:hep-th/0203116}, and, more recently, in \cite{Buividovich:22:2}. Truncation of the supersymmetric Hamiltonian (\ref{HS}) in the basis of harmonic oscillator states was considered for the first time in \cite{Wosiek:hep-th/0203116}. Unboundedness of classical trajectories and the coexistence of discrete and continuous spectra for the Hamiltonian (\ref{HS}) were demonstrated in \cite{Korcyl:hep-th/0610105}. The work \cite{Korcyl:hep-th/0610105} also presented a detailed analysis of the symmetry groups $C_{4v}$ and $D_{4d}$ of the bosonic and the supersymmetric Hamiltonians (\ref{HB}) and (\ref{HS}). Exact analytic results for the spectrum of supersymmetric model with $\mathcal{N} = 2$ supersymmetries were presented in \cite{Korcyl:0911.2152,Korcyl:1008.2975,Korcyl:1101.0591,Korcyl:1101.0668}.

In a recent work \cite{Buividovich:22:2} we considered the OTOCs for both the supersymmetric and the bosonic Hamiltonians (\ref{HS}) and (\ref{HB}) and estimated the quantum Lyapunov exponents. Like in \cite{Wosiek:hep-th/0203116,Korcyl:hep-th/0610105,Rinaldi:2108.02942}, we used numerical exact diagonalization in the truncated basis of harmonic oscillator states with level numbers $k_1$, $k_2$ (for coordinates $\hat{x}_1$ and $\hat{x}_2$), using only the states that satisfy $k_1 + k_2 < 2 M$. We found that supersymmetry results in a non-vanishing quantum Lyapunov exponent that scales as the first power of temperature $T$ down to the lowest temperatures. On the other hand, for the purely bosonic Hamiltonian (\ref{HB}) the Lyapunov exponent vanishes at a temperature $T \sim 1$. Therefore the quantum Lyapunov exponents for both Hamiltonians are consistent with the MSS bound, but only the supersymmetric model is chaotic at all temperatures. Quite unexpectedly, it turned out that at high temperatures the quantum Lyapunov exponents agree with their classical counterparts only for the supersymmetric system. We also found that the eigenstates of the supersymmetric Hamiltonian have a very regular low-dimensional structure at low energies. Nevertheless, the low-energy states maintain the growth of out-of-time-order correlators. At high temperatures or energies, both the bosonic and the supersymmetric Hamiltonians are chaotic, with energy spectra exhibiting the universal random-matrix-type level spacing fluctuations.

\begin{figure*}[h!tpb]
  \centering
  \includegraphics[width=0.49\textwidth]{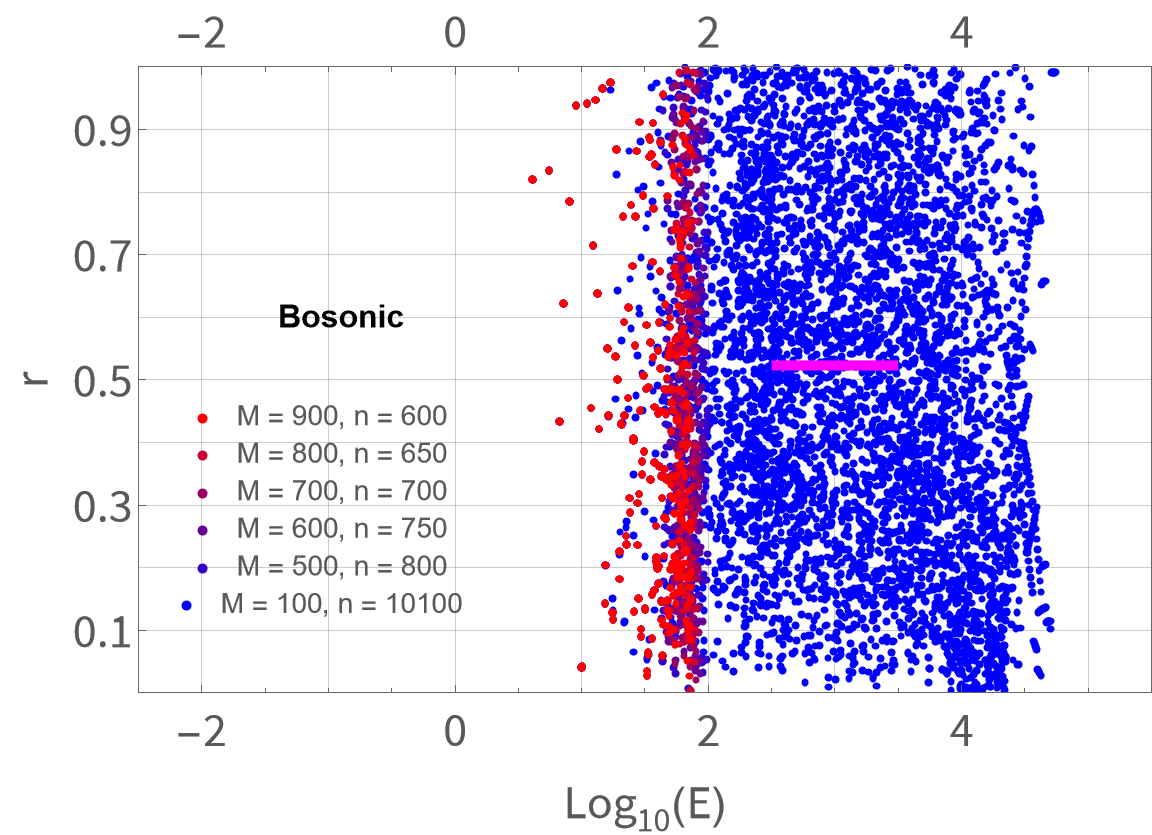}
  \includegraphics[width=0.49\textwidth]{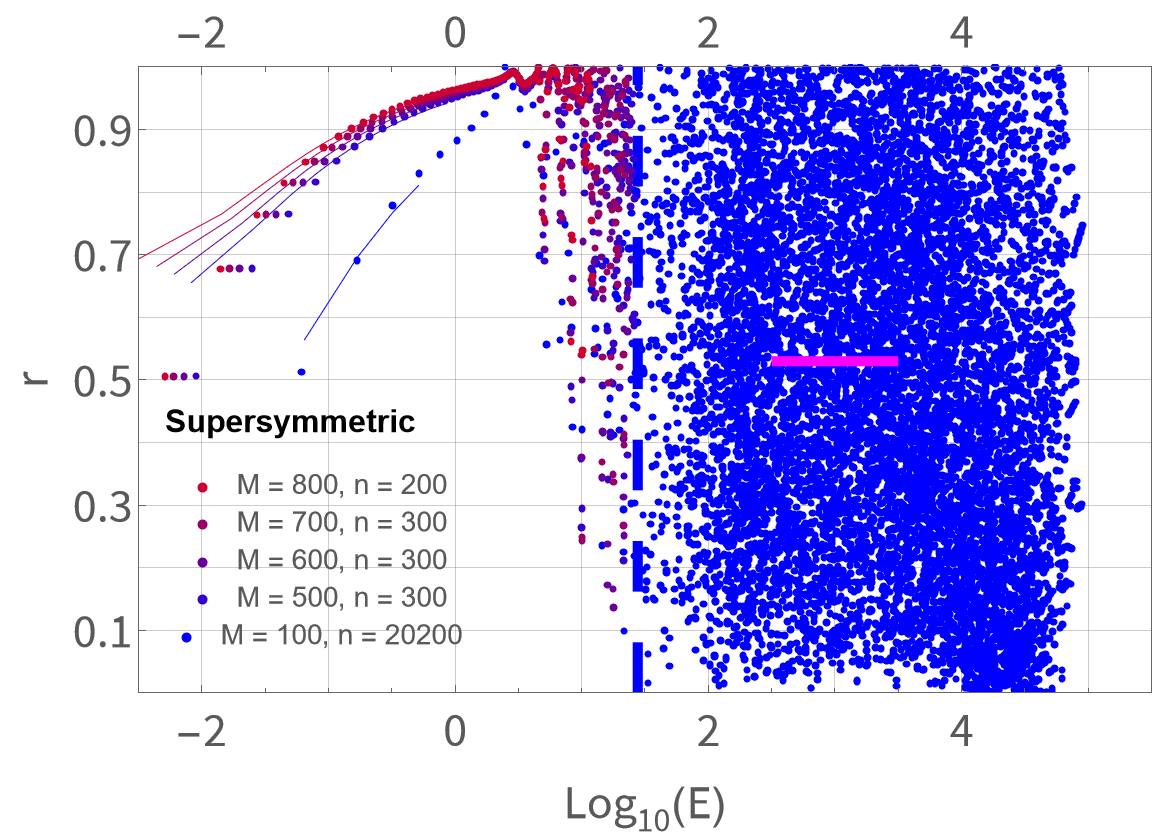}\\
  \caption{$r$-ratios $r_i$, defined in (\ref{r_ratio_def}), versus the energy level $E_i$. \textbf{On the right:} for the bosonic model (\ref{HB}). \textbf{On the left:} for the supersymmetric model (\ref{HS}). Only energy levels that transform under non-Abelian irreps $E_0$ of $C_{4v}$ and $E_1$ of $D_{4d}$ are analyzed. Solid magenta lines denote $r$-ratios averaged over all energy levels within the interval marked by the width of the lines. Vertical dashed lines show the energy after which the ordering of irreps becomes irregular.}
  \label{fig:r_ratios}
\end{figure*}

The transition between the regular low-energy eigenstates and the chaotic high-energy bulk of the spectrum appears to be quite sharp, even though the number of the degrees of freedom in our system is small. We conjecture that this transition might be similar to the transition between the $D0$-brane regime and the Schwarzschild black hole/$M$-theory regime in the BFSS matrix model, discussed in \cite{Maldacena:hep-th/9802042} and studied numerically in \cite{Hanada:2110.01312}. This transition becomes most obvious if we consider the $r$-ratios
\begin{eqnarray}
\label{r_ratio_def}
 r_i  = \frac{\min\lr{\Delta E_{i-1}, \Delta E_i}}{\max\lr{\Delta E_{i-1}, \Delta E_i}} ,
\end{eqnarray}
where $\Delta E_i = E_{i+1} - E_i$ is the spacing between the adjacent energy levels. The $r$-ratios are sensitive to statistical fluctuations in the energy spectrum and can be used to distinguish random-matrix-type statistics with repulsion between energy levels from Poisson statistics \cite{Huse:cond-mat/0610854,Luitz:1411.0660,Tezuka:1801.03204,Tezuka:2005.12809}. For systems with global symmetries, the $r$-ratios should be calculated for a subset of eigenstates that transform under the same irreducible representation (irrep) of the symmetry group. In our case, the symmetry group of the bosonic model is $C_{4v}$ with 5 irreps. The symmetry group of the supersymmetric model is $D_{4d}$ with 7 irreps \cite{Korcyl:hep-th/0610105,Buividovich:22:2}.

On Fig.~\ref{fig:r_ratios}, taken from our work \cite{Buividovich:22:2}, we compare the energy dependence of the $r$-ratios for the bosonic Hamiltonian (\ref{HB}) and for the supersymmetric Hamiltonian (\ref{HS}). One can see that at sufficiently high energies, the $r$-ratios are exhibiting very irregular behavior, filling almost uniformly all values between $0$ and $1$. With sufficiently many energy levels, this allows us to perform a statistical average over some interval of energies. For the high-energy spectra of both the supersymmetric and the bosonic model, such averaging yields values that are very close to the universal value $\bar{r}_{GOE} = 0.53$ for the Gaussian Orthogonal Ensemble \footnote{Since both the bosonic and the supersymmetric Hamiltonians are manifestly real \cite{Buividovich:22:2}, the relevant ensemble here is the Gaussian Orthogonal Ensemble.}. For the supersymmetric Hamiltonian, however, the $r$-ratios for the low-energy eigenstates behave in a completely regular way and do not exhibit any quasi-statistical fluctuations. They are consistent with the phenomenological expression $E_i = A\lr{M} + B\lr{M} \, i + C\lr{M} \, i^2$, where $i$ is the serial number of the energy level for a given irrep. The transition between the regular and the chaotic behaviors of $r$-ratios appears to be quite sharp and happens around $E = 10$, independently of the value of the regularization parameter $M$. This transition appears to be a distinctive feature of our supersymmetric model. We also explicitly checked that the regular behavior of $r$-ratios is absent in the spectrum of the SYK model\footnote{We thank Masaki Tezuka for kindly providing numerical data for the energy spectrum of the SYK model that was used in \cite{Hanada:1611.04650}.}.

In these Proceedings, we study how the transition between the low-energy regular eigenstates and the high-energy chaotic eigenstates of the supersymmetric Hamiltonian (\ref{HS}) manifests itself in quantities other than $r$-ratios. Specifically, we will consider the thermodynamic equation of state (Section~\ref{sec:EoS}), the real-time spectral form-factors (Section~\ref{sec:SPF}), and the entanglement Hamiltonians and entanglement entropy of excited states (Section~\ref{sec:entanglement}). All these quantities can be also studied within the framework of holographic AdS/CFT correspondence. We hope that our results might shed light on the microscopic mechanism of the transition between the $D0$-brane regime and the Schwarzschild black hole/$M$-theory regime in the BFSS model \cite{Maldacena:hep-th/9802042,Hanada:2110.01312}. As in our work \cite{Buividovich:22:2}, we will highlight the differences in the behavior of the supersymmetric and the bosonic Hamiltonians throughout these Proceedings.

\begin{figure*}[h!tpb]
  \centering
  \includegraphics[width=0.49\textwidth]{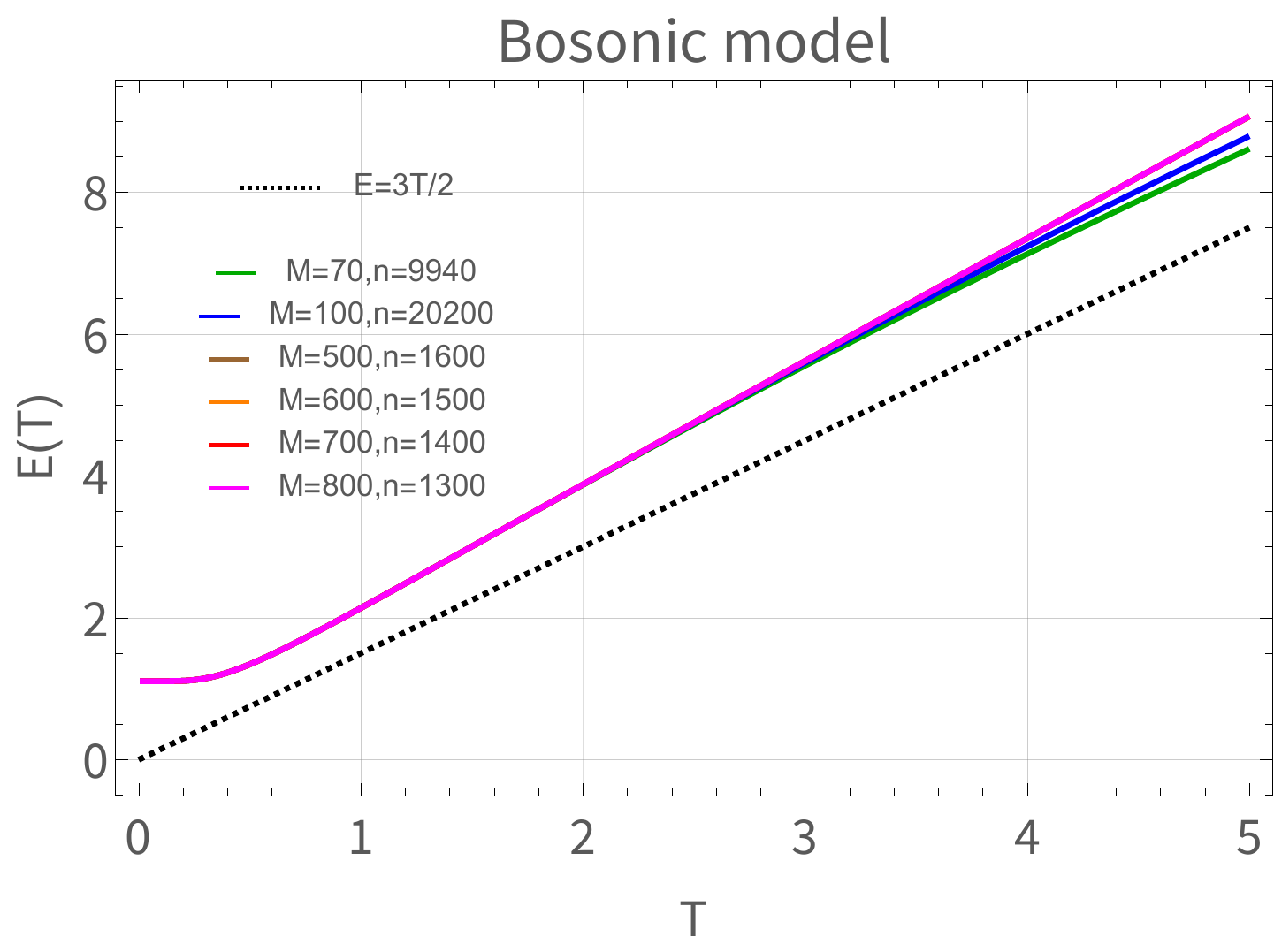}
  \includegraphics[width=0.49\textwidth]{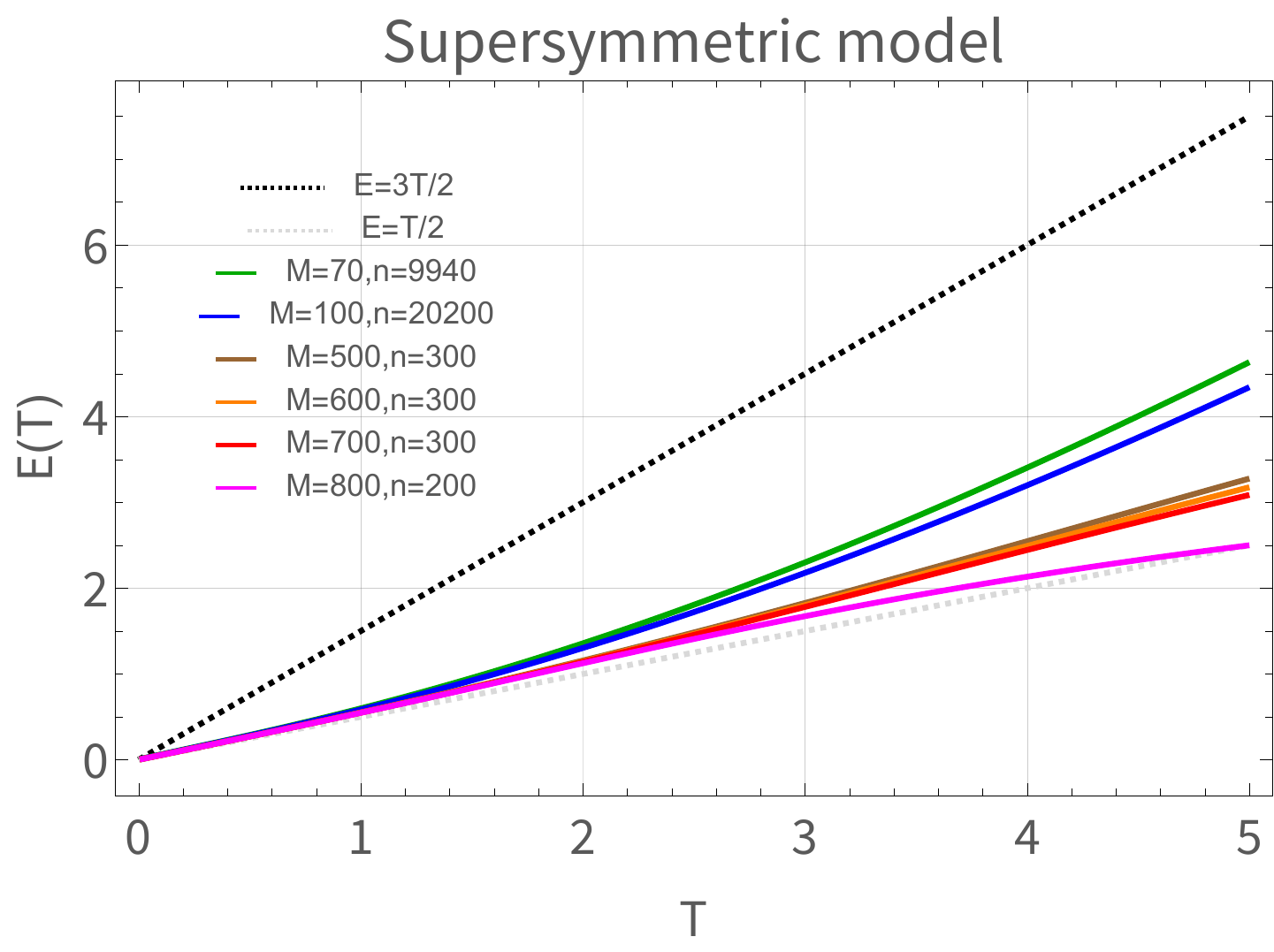}\\
  \includegraphics[width=0.49\textwidth]{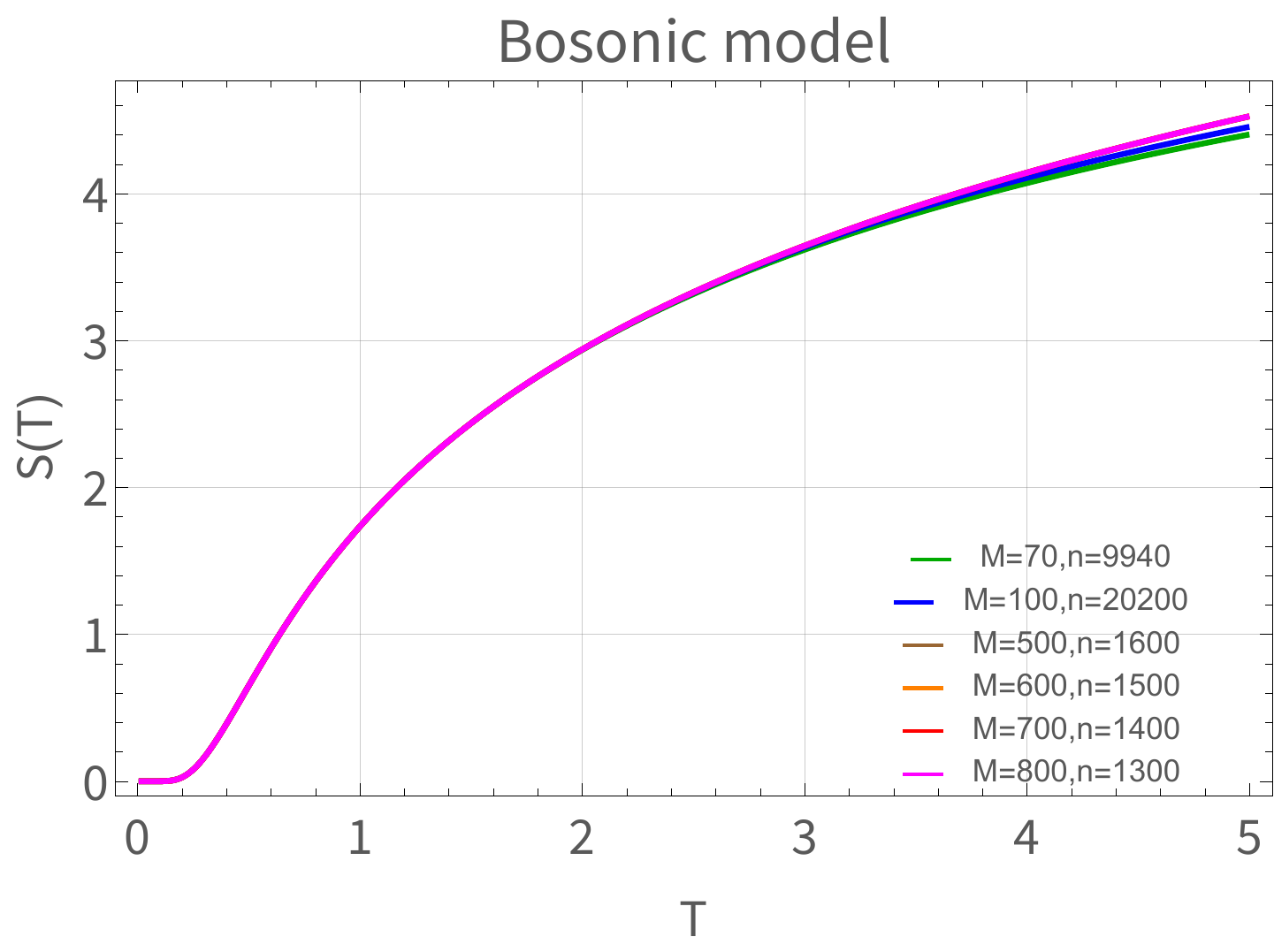}
  \includegraphics[width=0.49\textwidth]{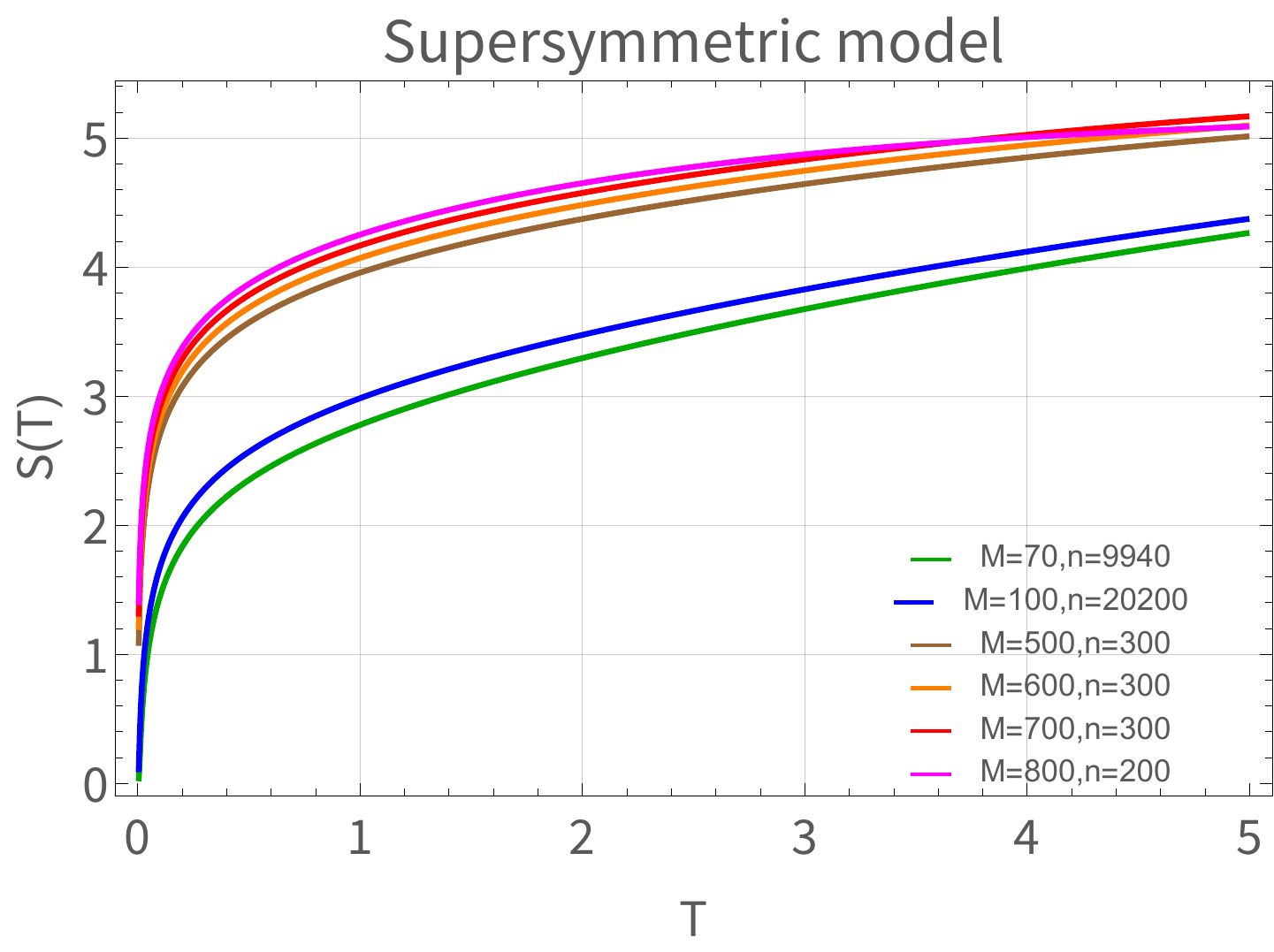}\\
  \includegraphics[width=0.49\textwidth]{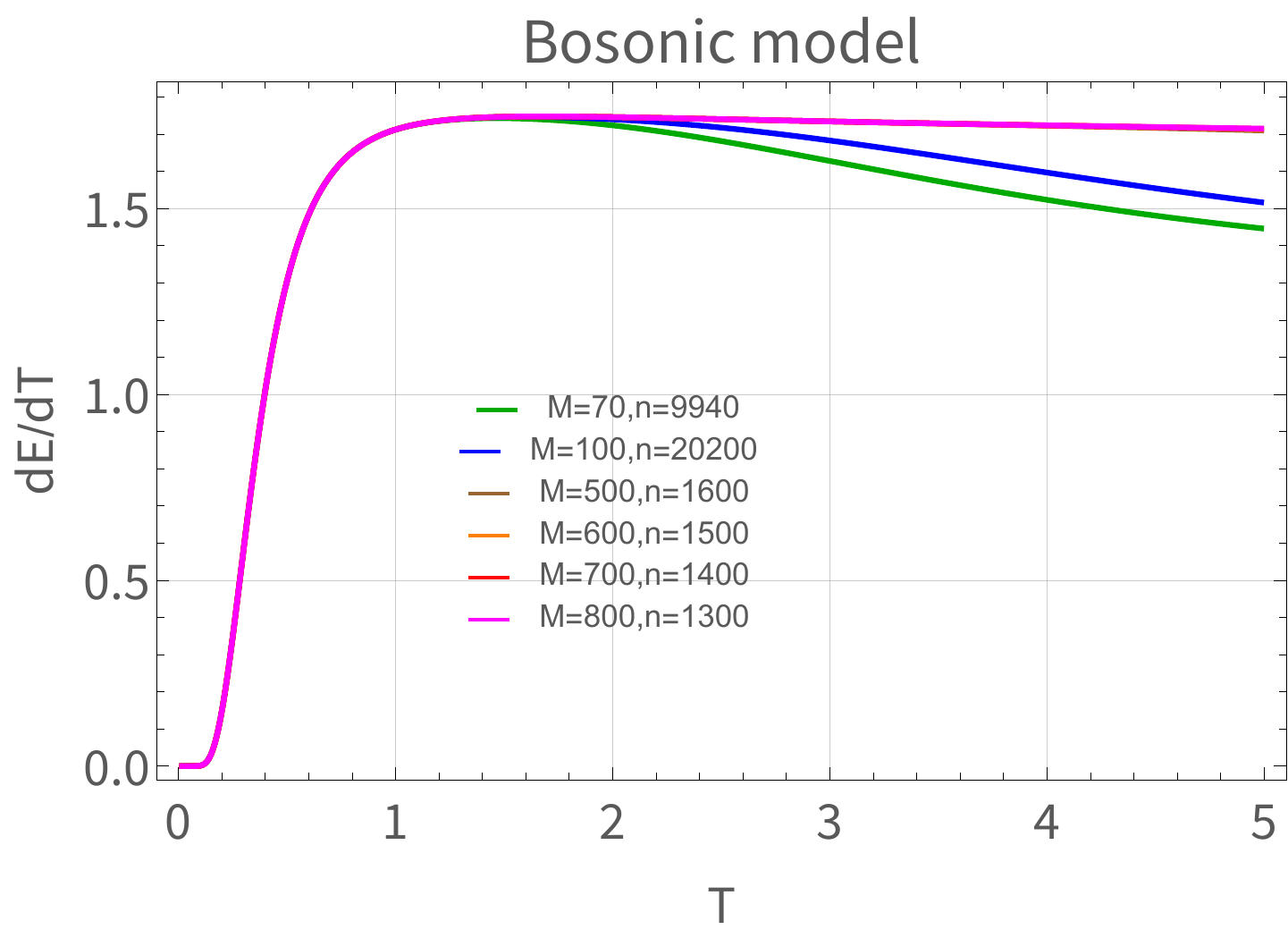}
  \includegraphics[width=0.49\textwidth]{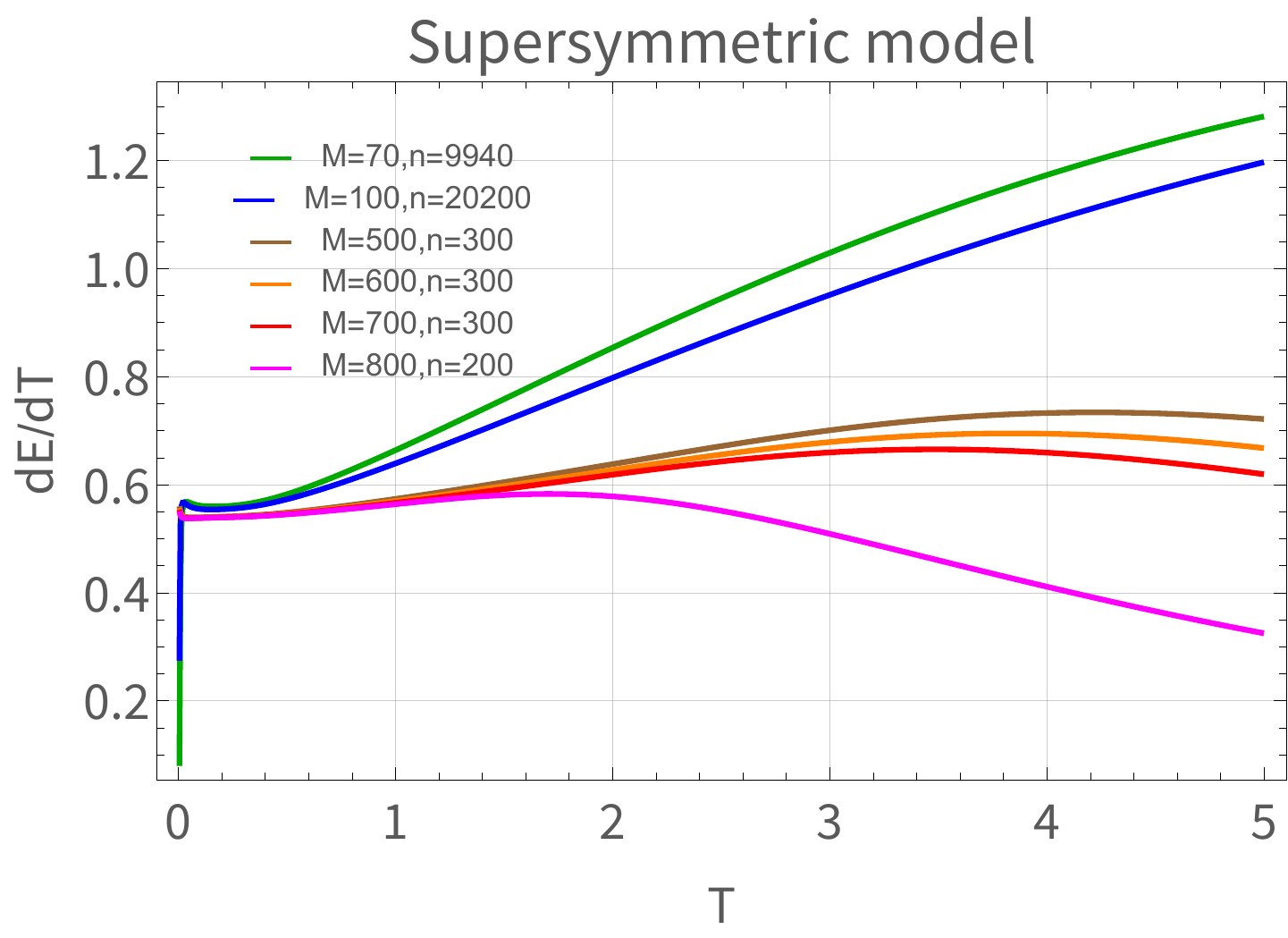}\\
  \caption{Mean energy $E\lr{T}$ (\textbf{top row}), thermodynamic entropy $S\lr{T}$ (\textbf{middle row}), and heat capacity $d E\lr{T}/d T$ (\textbf{bottom row}) as functions of the temperature $T$ for the bosonic model (\ref{HB}) (\textbf{on the left}) and for the supersymmetric model (\ref{HS}) (\textbf{on the right}). $M$ in the plot legend is the Hilbert space truncation parameter, and $n$ is the number of lowest energy levels used to calculate the equation of state.}\label{fig:EoS}
\end{figure*}

\section{Equation of state}
\label{sec:EoS}

Let us first consider the thermodynamic equation of state for both the bosonic and the supersymmetric Hamiltonians (\ref{HB}) and (\ref{HS}). We use numerical results for the energy levels $E_i$ to calculate the mean energy $E\lr{T}$, the thermodynamic entropy $S\lr{T}$, and the heat capacity $d E\lr{T}/dT$:
\begin{eqnarray}
\label{EvsT}
 E\lr{T} = Z^{-1}\lr{T} \, \sum\limits_i E_i \, \exp\lr{-E_i/T} ,
 \quad
 Z\lr{T} = \sum\limits_i \exp\lr{-E_i/T} ,
 \\
 \label{SvsT}
 S\lr{T} = \ln\lr{Z} + T^{-1} \, E\lr{T} ,
 \\
 \label{dEvsdT}
 \frac{d E\lr{T}}{d T} = \frac{1}{Z\lr{T} \, T^2} \, \sum\limits_i \lr{E_i - E\lr{T}}^2 \, \exp\lr{-E_i/T} .
\end{eqnarray}

It is useful to compare the numerical results with the classical equation of state for the bosonic Hamiltonian (which should also be the classical limit of the supersymmetric system). To obtain the classical equation of state, we consider the classical thermodynamic partition function
\begin{eqnarray}
\label{classical_partition0}
 Z_{cl}\lr{T}
 =
 \int d p_1 \, d p_2 \exp\lr{-\frac{p_1^2 + p_2^2}{T}}
 \,
 \int d x_1 \, d x_2 \exp\lr{-\frac{x_1^2 x_2^2}{T} } .
\end{eqnarray}
We now perform integrations over the momenta $p_1$ and $p_2$, and express the integration over $x_1$ and $x_2$ in terms of the new variables $r$ and $\phi$ as $x_1 = r \, e^{\phi}$, $x_2 = r \, e^{-\phi}$. This results in
\begin{eqnarray}
\label{classical_partition1}
 Z_{cl}\lr{T}
 =
 2 \, \pi \, T \, \int\limits_{-\infty}^{+\infty} d\phi \int\limits_0^{+\infty} dr \, r \, \exp\lr{-\frac{r^4}{T} } .
\end{eqnarray}
We see that integration over the ``hyperbolic angle'' $\phi$ completely factors out. Changing the integration variable from $r$ to $u = r/T^{1/4}$, we find the explicit dependence of $Z_{cl}\lr{T}$ on the temperature $T$ up to an overall normalization factor $\mathcal{N}$, which also absorbs any divergences due to infinite range of integration over $\phi$:
\begin{eqnarray}
\label{classical_partition2}
 Z_{cl}\lr{T}
 =
 \mathcal{N} \, T^{3/2} .
\end{eqnarray}
From this, we immediately find the classical equation of state in which the multiplicative divergence of $Z_{cl}\lr{T}$ cancels out:
\begin{eqnarray}
\label{EoS_classical}
 E_{cl}\lr{T} = T^2 \frac{\partial \log\lr{Z_{cl}\lr{T}}}{\partial T} = \frac{3}{2} \, T .
\end{eqnarray}

We show our numerical results for the equation of state on Fig.~\ref{fig:EoS}, comparing the cases of the bosonic Hamiltonian (\ref{HB}) (on the left) and the supersymmetric Hamiltonian (\ref{HS}) (on the right). We also show the classical equation of state (\ref{EoS_classical}) on the plots of the energy $E\lr{T}$. We see that the equation of state for the bosonic model has a typical behavior for a gapped system: $E\lr{T}$ tends to a finite value at $T = 0$, while the entropy and the heat capacity approach zero. At higher energies, the equation of state becomes similar to the classical one. Effects of Hilbert space truncation appear to be small for the bosonic case.

On the other hand, for the supersymmetric model the energy $E\lr{T}$ goes to zero at $T = 0$. The heat capacity (or, equivalently, the slope of the function $E\lr{T}$) is again close to the classical value $\frac{d E_{cl}\lr{T}}{d T} = \frac{3}{2}$ (\ref{EoS_classical}) at high energies. At low energies, it appears to be very close to the value $\frac{d E_{1D}\lr{T}}{d T} = \frac{1}{2}$ for a one-dimensional ideal gas, in full accordance with the nearly one-dimensional structure of the corresponding wave functions \cite{Buividovich:22:2}.

The effects of Hilbert space truncation are quite strong for the supersymmetric Hamiltonian, as can be expected for a system with a continuous spectrum. In particular, the high-temperature results are completely unreliable at large values of $M$, where we use $n = \mathcal{O}\lr{100}$ lowest eigenstates to calculate thermodynamic quantities. At sufficiently low temperatures, however, all results converge and are numerically reliable.

Overall, we do not see any signatures of thermodynamic singularity at intermediate temperatures in any thermodynamic quantities. This suggests that, at least for our low-dimensional models, the sharp transition between the regular and chaotic behavior of Hamiltonian eigenstates, as seen on the right plot on Fig.~\ref{fig:r_ratios}, is not manifesting itself in a sharp change of any thermodynamic quantities. Of course, there is still a chance that such thermodynamic singularities can still appear in the full BFSS model, especially in the large-$N$ limit.

\section{Spectral form-factors}
\label{sec:SPF}

\begin{figure}[h!tpb]
  \centering
  \includegraphics[width=0.49\textwidth]{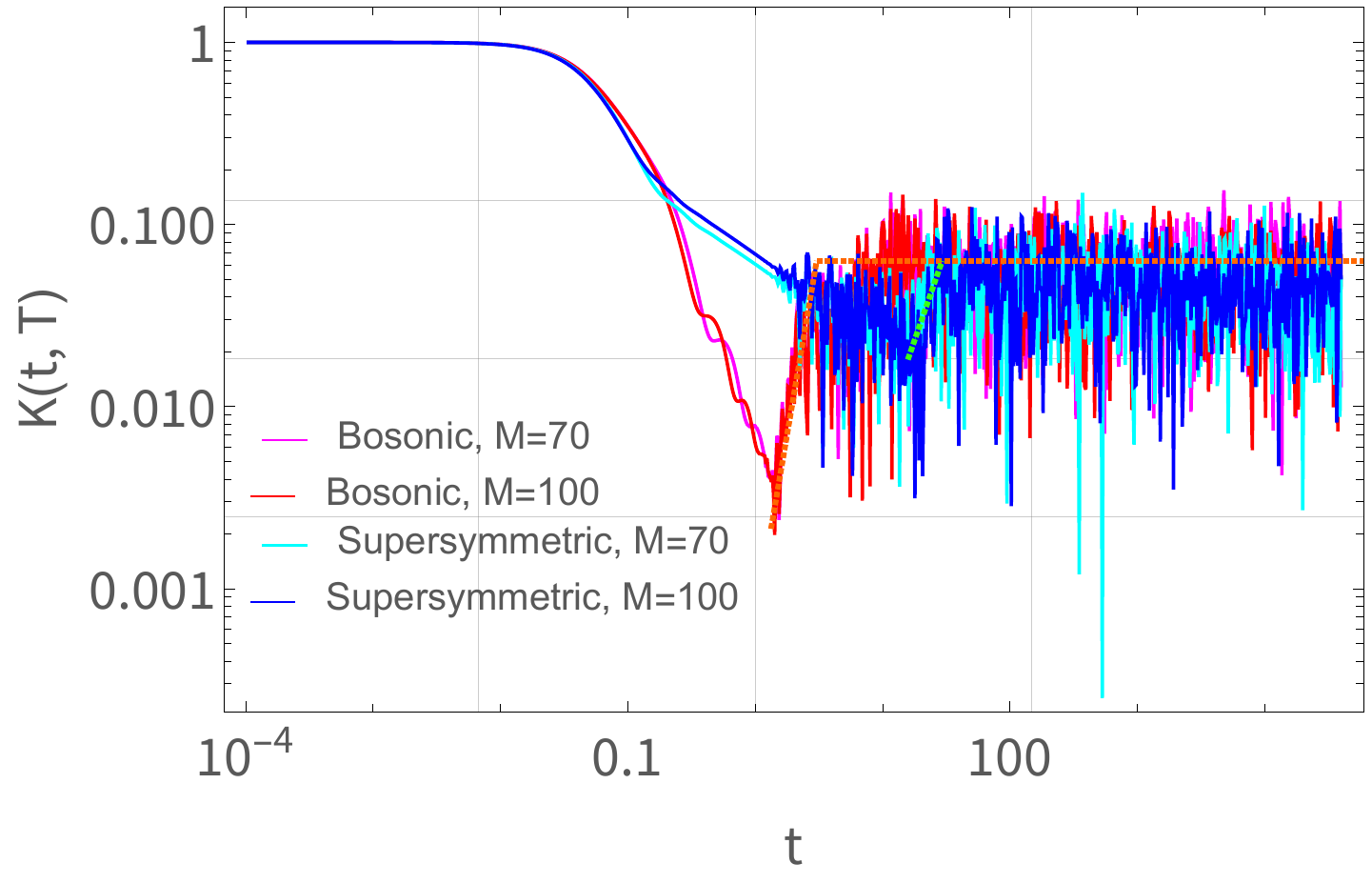}\\
  \caption{Real-time spectral form-factors (\ref{SPF_def}) as functions of the evolution time $t$ for the bosonic model (\ref{HB}) (\textbf{on the left}) and for the supersymmetric model (\ref{HS}) (\textbf{on the right}). The temperature is $T = 20$. Dashed lines visually highlight the approximate ``ramp'' behavior.}\label{fig:SPF}
\end{figure}

Real-time spectral form-factors
\begin{eqnarray}
\label{SPF_def}
 K\lr{t} = Z^{-1}\lr{T} \, \left| \tr \exp\lr{i \hat{H} t - \hat{H}/T} \right|
\end{eqnarray}
provide an alternative way to diagnose quantum chaos using real-time quantities. Here we use the finite-temperature partition function $Z\lr{T}$, defined in (\ref{EvsT}), to impose the normalization $K\lr{0} = 1$. While spectral form-factors do not have such a clear physical interpretation as out-of-time-order correlators, they are technically much easier to calculate. From theoretical point of view, one can consider the infinite-temperature limit of $K\lr{t}$, but in practical calculations a cut-off on higher energy levels is often required to remove noise \cite{Hanada:1803.08050} and turn $K\lr{t}$ into a relatively smooth function. Introducing finite temperature is one possible way to introduce such a cut-off.

After initial quick decay, spectral form-factors for chaotic systems with universal random-matrix-type correlations between energy levels described by a Gaussian Unitary Ensemble exhibit a period of linear-in-time growth $K\lr{t} \sim t$, followed by saturation \cite{Yoshida:1706.05400,Hanada:1803.08050}. This time dependence looks like a characteristic ramp on the log-log-scale plot of $K\lr{t}$. For the Gaussian Orthogonal Ensemble (GOE), the ``cusp'' of the ramp is smoothed out (see Section 3.2.2 in \cite{Liu:1806.05316}), but otherwise does not significantly change its shape in comparison to the GUE result.

We show our results for the real-time spectral form-factors $K\lr{t}$ of the bosonic and the supersymmetric Hamiltonians at a high temperature $T = 20$ on Fig.~\ref{fig:SPF}. As expected, for both the bosonic and the supersymmetric cases we see a period of approximately linear growth after initial fast decay, which is followed by saturation at late times. Numerical noise, however, obstructs a clear identification of the ramp and its slope. Dashed lines on Fig.~\ref{fig:SPF} are just intended to guide the eye. Presumably, we need larger values of the cutoff $M$ wih more eigenstates $n$ to produce better plots for spectral form-factors. It is also possible that spectral form-factors simply work better for systems with quenched disorder. With the available data, further decreasing the temperature $T$ does not remove the noise, but merely increases the typical scale of fluctuations, which obstructs the ramp identification even more. So our value $T = 0.05$ is close to the optimal value that allows to see something like a ramp. It is also interesting to note that before the onset of the ramp behavior, spectral form-factors for both Hamiltonians (\ref{HB}) and (\ref{HS}) exhibit a period of clear power-law decay, which looks like a linear dependence of $\log\lr{K\lr{t}}$ on $\log\lr{t}$ on our log-log scale plot on Fig.~\ref{fig:SPF}. The decay/growth rates at early times and in the ramp region are, however, clearly different for the bosonic and for the supersymmetric Hamiltonian, even though they both should belong to the same GOE universality class.

\begin{figure*}[h!tpb]
  \centering
  \includegraphics[width=0.49\textwidth]{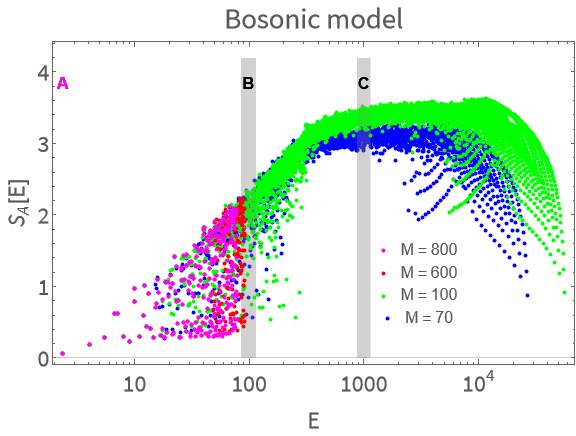}
  \includegraphics[width=0.49\textwidth]{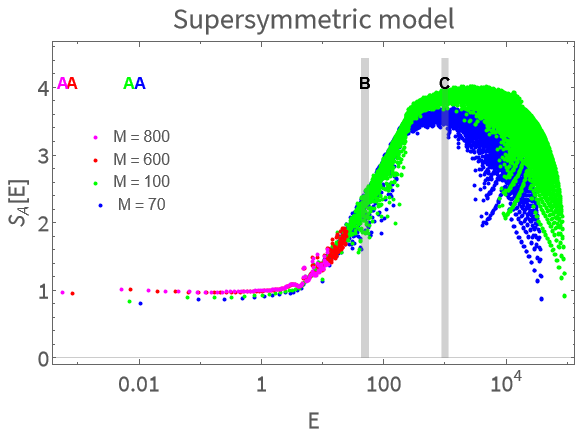}
  \caption{Energy dependence of the eigenstate entanglement entropy $S_A\lr{E}$, where the observable subsystem is one of the bosonic degrees of freedom. \textbf{On the right:} for the bosonic model (\ref{HB}). \textbf{On the left:} for the supersymmetric model (\ref{HS}). Opaque vertical bars marked with letters denote the energy intervals for which we show the spectra of the entanglement Hamiltonian on Fig.~\ref{fig:entanglement_spectra}.}
  \label{fig:EEvsE}
\end{figure*}

\begin{figure}[h!tpb]
  \centering
  \includegraphics[width=0.49\textwidth]{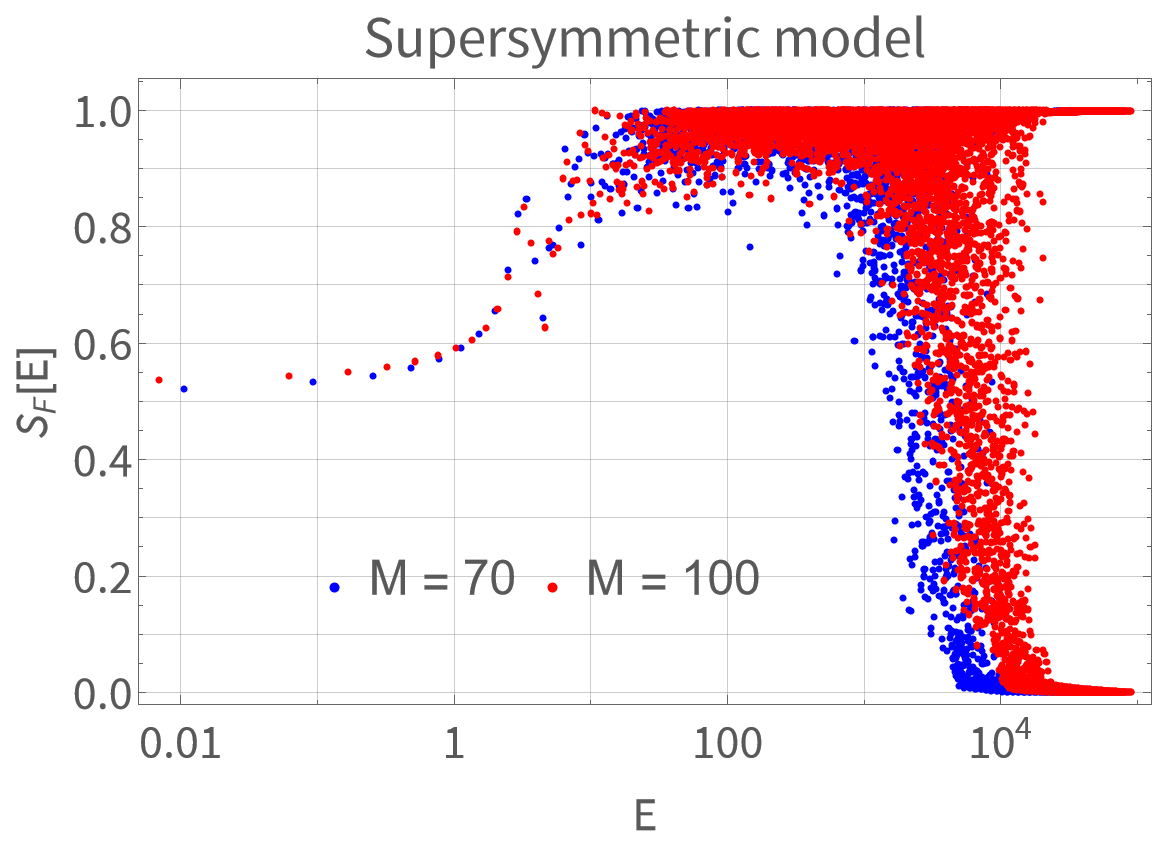}
  \caption{Energy dependence of the entanglement entropy of the fermionic degree of freedom for excited states.}
  \label{fig:EEvsE_fermionic}
\end{figure}

\section{Eigenstate entanglement entropy}
\label{sec:entanglement}

Ground-state entanglement entropy has attracted enormous attention in quantum field theory and in quantum many-body physics as a universal probe of quantum transitions and topological order that is independent of any order parameters. Entanglement properties of higher eigenstates provide even more information about many-body quantum systems \cite{Alba:0909.1999}. Eigenstate entanglement has also proven useful in the context of many-body quantum chaos, in particular, as a probe of Eigenstate Thermalization Hypothesis (ETH). In systems where ETH holds, entanglement entropy of excited states leads to emergent thermodynamics for sub-systems of a large system \cite{Deutsch:0911.0056,Polkovnikov:1202.4764,Polkovnikov:1509.06411,Huang:1708.08607}. For an observer that has only access to a relatively small part of the whole system, typical eigenstates in the chaotic bulk of the spectrum look like thermal states. All the information that is available by accessing only a part $A$ of the whole system that is in an eigenstate $\ket{\Psi}$ on a Hamiltonian $\hat{H}$ is encoded in its reduced density matrix $\hat{\rho}_A$:
\begin{eqnarray}
\label{rho_def}
 \hat{\rho}_A = \tr_B \ket{\Psi} \bra{\Psi} \equiv e^{-\hat{H}_A}  ,
\end{eqnarray}
where $\tr_B\lr{ \ldots }$ denotes a trace over all degrees of freedom that do not belong to subsystem $A$, collectively denoted as $B$. Sometimes one also writes the reduced density matrix in terms of an entanglement Hamiltonian $\hat{H}_A$, that is often referred to as a ``modular Hamiltonian'' in the literature on holographic duality. Entanglement entropy of a state $\ket{\Psi}$ is defined as the von Neumann entropy of the reduced density matrix $\hat{\rho}_A$:
\begin{eqnarray}
\label{EE_def}
 S_A = - \tr_A \lr{ \hat{\rho}_A \, \log\lr{\hat{\rho}_A} } .
\end{eqnarray}
Plotting the entanglement entropy of eigenstates as a function of the energy of the state, we can obtain the entanglement ``equation of state'' $S_A\lr{E}$ , and even define the ``entanglement temperature'' $T_A^{-1} = \frac{d S_A\lr{E}}{d E}$ \cite{Takayanagi:1212.1164,Naseh:1305.2728}.

In the context of holographic duality, ground-state entanglement entropy of conformal field theories is interpreted as a minimal area of the surface in a holographic dual space-time that spans on the boundary of a subsystem $A$ \cite{Ryu:hep-th/0603001}. Likewise, entanglement entropy of higher eigenstates is dual to minimal surfaces in space-times that are perturbations of the ground-state dual space-time \cite{Wong:1305.3291}. In this respect, eigenstate entanglement entropy allows to trace how the dual space-time changes upon injecting more energy into the system in an adiabatic way.

Although our Hamiltonians $\hat{H}_B$ and $\hat{H}_S$ have just a few degrees of freedom, we can still calculate the entanglement entropies and the entanglement Hamiltonians. For the bosonic system with the Hamiltonian (\ref{HB}), the only possibility is to consider the entanglement between the Hilbert spaces associated with the two bosonic degrees of freedom $\hat{x}_1$ and $\hat{x}_2$. Correspondingly, we obtain the reduced density matrix by tracing out $x_2$:
\begin{eqnarray}
\label{rho_bosonic_def}
 \rho_A\lr{x_1, x_1'} = \int \limits_{-\infty}^{+\infty} d x_2 \, \Psi\lr{x_1, x_2} \, \bar{\Psi}\lr{x_1', x_2} ,
\end{eqnarray}
where $\Psi\lr{x_1, x_2}$ is the wave function of the eigenstate $\ket{\Psi}$. For the supersymmetric system with the Hamiltonian (\ref{HS}), we have two options. The first is to consider $x_1$ as a subsystem and to trace out $x_2$ and the fermionic degree of freedom. The second one is to consider the fermionic degree of freedom as a subsystem, and to trace out both bosonic degrees of freedom. Correspondingly, we obtain two different density matrices, which we denote as $\hat{\rho}_A$ and $\hat{\rho}_F$, where the subscript ``F'' stands for ``fermionic'':
\begin{eqnarray}
\label{rho_supersymmetric_bosonic_def}
 \rho_A\lr{x_1, x_1'} = \int \limits_{-\infty}^{+\infty} d x_2 \, \sum\limits_{\alpha} \, \Psi_{\alpha}\lr{x_1, x_2} \, \bar{\Psi}_{\alpha}\lr{x_1', x_2} ,
\\
\label{rho_supersymmetric_fermionic_def}
 \lr{\rho_F}_{\alpha\beta} = \int \limits_{-\infty}^{+\infty} d x_1 \, d x_2 \, \Psi_{\alpha}\lr{x_1, x_2} \, \bar{\Psi}_{\beta}\lr{x_1, x_2} .
\end{eqnarray}
The indices $\alpha, \beta = 1, 2$ correspond to the fermionic part of the Hilbert space, acted upon by the Pauli matrices in (\ref{HS}). The corresponding entanglement entropies, defined according to (\ref{EE_def}), are denoted as $S_A$ and $S_F$. With our truncations of the Hilbert space, the density matrix $\rho_A$ is equivalent to an $M \times M$ matrix for both the bosonic and the supersymmetric systems.

On Fig.~\ref{fig:EEvsE} we show the dependence of the eigenstate entanglement entropies on the eigenstate energy $E$ for the bosonic and the supersymmetric systems, where the subsystem $A$ is the bosonic degree of freedom $x_1$. Again, on all plots we select only energy levels that transform under a non-Abelian irrep $\mathcal{E}_0$ (in the bosonic case) or $\mathcal{E}_1$ (in the supersymmetric case), see Appendix~A in \cite{Buividovich:22:2} for more details. For other irreps, the situations is qualitatively similar.

For the bosonic Hamiltonian, the entanglement entropy $S_A$ grows as approximately $\log\lr{E}$ for $E \lesssim 10^2$. For $E \gtrsim 10^2$, $S_A$ reaches a plateau with a height of approximately $S_A\lr{E \gtrsim 10^2} \approx 3.5$. This value is close to, but somewhat smaller than the Page value $S_{Page} = \log_2\lr{M} - \frac{1}{2}$ of the entanglement entropy of a random state distributed uniformly over the entire Hilbert space. For $M=70$, $S_{Page} = 3.7$. For $M = 100$, $S_{Page} = 4.1$. The saturation of $S_A\lr{E}$ happens at approximately the same energy at which the $r$-ratio values on the left plot on Fig.~\ref{fig:r_ratios} fill the entire interval $r \in \lrs{0, 1}$. The saturation of entanglement entropy for the high-energy bulk of the spectrum supports the validity of the ETH for these eigenstates.

For the supersymmetric system (\ref{HS}), the entanglement entropy $S_A\lr{E}$ practically does not depend on the energy $E$ in the low-energy part of the spectrum that corresponds to regularly behaved $r$-ratios on the right plot on Fig.~\ref{fig:r_ratios}. The value of $S_A$ for this part of the spectrum is also quite small. For energies $E \gtrsim 10^1$, where the $r$-ratios start exhibiting random fluctuations, the entanglement entropy also starts growing in a way similar to the one in the bosonic model. This growth saturates at $E \gtrsim 5 \cdot 10^2$, where also the $r$-ratios on the right plot on Fig.~\ref{fig:r_ratios} are fully random. The saturation value is also reasonably close to the Page limit $S_{Page}$.

For completeness, on Fig.~\ref{fig:EEvsE_fermionic} we also show the energy dependence of the fermionic entanglement entropy $S_F$. It also has a weak energy dependence for the low-energy part of the spectrum, and gradually grows to the maximal possible value $S_F = \log\lr{2}$ towards higher energies.

\begin{figure*}[h!tpb]
  \centering
  \includegraphics[width=0.49\textwidth]{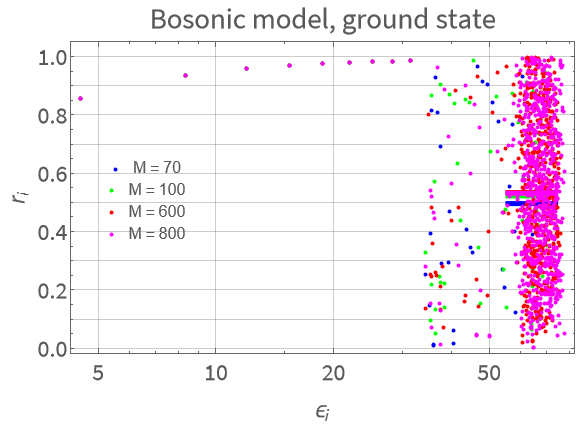}
  \includegraphics[width=0.49\textwidth]{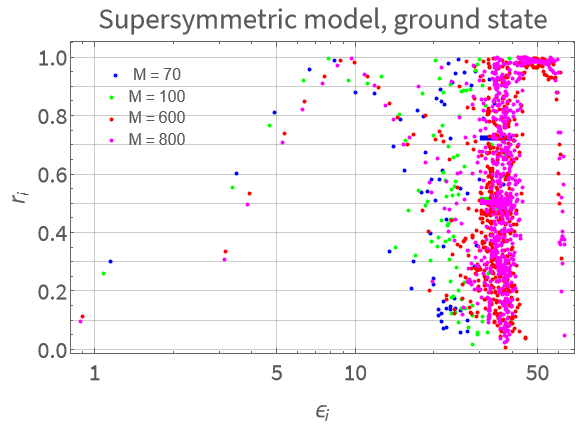}\\
  \includegraphics[width=0.49\textwidth]{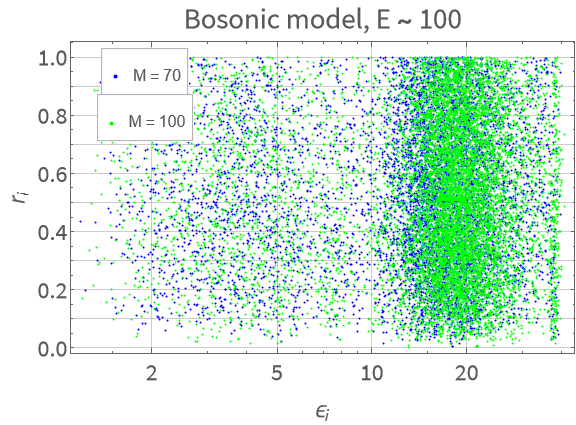}
  \includegraphics[width=0.49\textwidth]{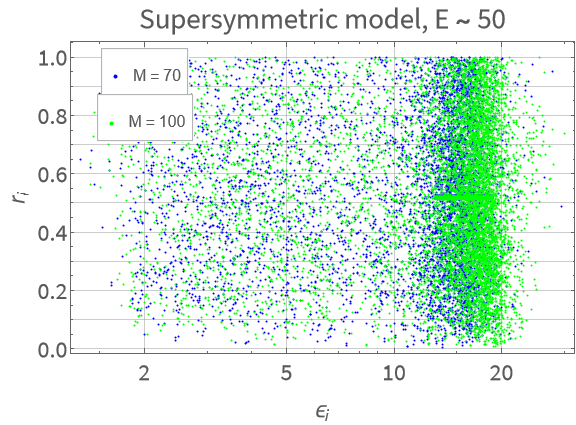}\\
  \includegraphics[width=0.49\textwidth]{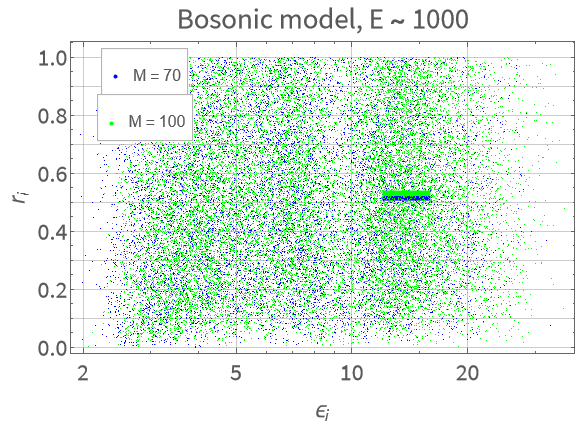}
  \includegraphics[width=0.49\textwidth]{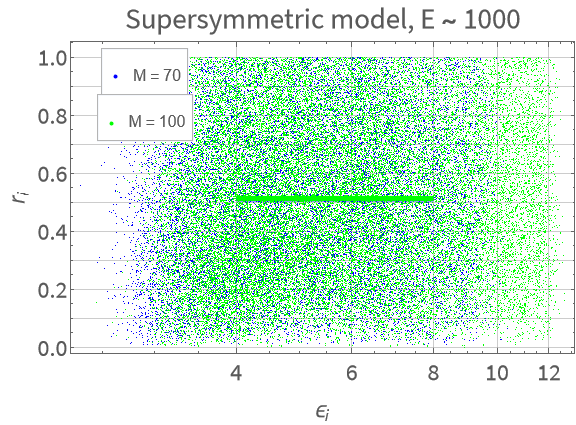}\\
  \caption{Entanglement spectra of excited states at different energies for the bosonic (\textbf{on the right}) and the supersymmetric (\textbf{on the left}) models. In the upper row, only the ground-state entanglement spectrum is shown. In the two lower rows, spectra of all eigenstates within the shaded bands on Fig.~\ref{fig:EEvsE} are combined together.}
  \label{fig:entanglement_spectra}
\end{figure*}

Finally, due to numerical exact diagonalization providing the full access to the wave functions of the system, we can also calculate the full spectrum of the entanglement Hamiltonian $\hat{H}_A$ in (\ref{rho_def}). Entanglement Hamiltonian is a much more detailed probe of the system than the entanglement entropy. It has been demonstrated to have many universal properties in the context of conformal field theories and holographic duality. For spherical entangled regions $A$ in conformal field theories, the entanglement Hamiltonian is related to a certain integral of the component $T_{00}$ of the energy-momentum tensor over the region $A$ \cite{Casini:1102.0440}. For two-dimensional conformal field theories, the spectrum of $\hat{\rho}_A$ is universal and only depends on the central charge \cite{Calabrese:0806.3059}.

For systems with quantum chaos, one can ask whether the spectrum of the entanglement Hamiltonian inherits statistical properties of the full spectrum of the system. This was demonstrated in \cite{Tomsovic:1807.00572} using perturbation theory combined with random matrix techniques. In these Proceedings we will numerically demonstrate that this statement is true for our Hamiltonian systems which feature quantum chaos.

To analyze our numerical data for the spectrum of entanglement Hamiltonian $\hat{H}_A$, we calculate the $r$-ratios (\ref{r_ratio_def}) for the eigenvalues $\epsilon_i$ of $\hat{H}_A$ (that are just minus the logs of the eigenvalues of $\hat{\rho}_A$), and plot them as functions of $\epsilon_i$ on Fig.~\ref{fig:entanglement_spectra}. In all cases, the subsystem $A$ is a coordinate $x_1$. We consider three different ranges of energy: first, in the top two plots on Fig.~\ref{fig:entanglement_spectra} we show the entanglement spectrum for the ground states of both the bosonic and the supersymmetric Hamiltonians. In the middle-row and lowest-row plots, we combine the entanglement spectra for all eigenstates within a finite-width range of energies, shown as shaded bands on Fig.~\ref{fig:EEvsE}. The middle-row plots correspond to the parts of the spectra of $\hat{H}_B$ and $\hat{H}_S$ where the entanglement entropy grows with energy. In the bottom-row plots, we also consider a range of energies for which the entanglement entropy saturates near the Page value.

Ground-state entanglement spectra appear to have quite a nontrivial structure for both bosonic and supersymmetric Hamiltonians. In the bosonic case, the $r$-ratios of $\epsilon_i$ stay very close to $r = 1$ and do not exhibit noticeable statistical fluctuations at small $\epsilon$. At higher $\epsilon$, they start fluctuating randomly and eventually fill up the entire range $r \in \lrs{0, 1}$. Averaging $r_i$ over a finite range of $\epsilon_i$, which is again shown on Fig.~\ref{fig:entanglement_spectra} as a solid horizontal line, we obtain the values that are close to the GOE random matrix ensemble result $\bar{r}_{GOE} = 0.53$.

In the supersymmetric case, the ground state apparently belongs to the family of very regular low-energy states. The behavior of $r$-ratios at low $\epsilon_i$ again appears to be regular, but not monotonous with respect to $\epsilon_i$. What is most nontrivial, however, is that at high values of $\epsilon$ the $r$-ratios also fluctuate randomly and fill up the entire range $r \in \lrs{0, 1}$. This implies that even though the ground state of our supersymmetric model (\ref{HS}) is deeply in the regime of regular, low-dimensional eigenstates, the information about the chaotic behavior in the high-energy bulk of the spectrum is somehow still encoded in its reduced density matrix.

Looking at the entanglement spectra of the higher-energy eigenstates (middle and bottom row on Fig.~\ref{fig:entanglement_spectra}), we note that the $r$-ratios are randomly fluctuating and filling the entire interval $r \in \lrs{0, 1}$ at more or less all values of $\epsilon$. In full agreement with the predictions of \cite{Tomsovic:1807.00572}, we find that mean values of $r$ averaged over a finite range of $\epsilon$ are very close to the GOE result $\bar{r}_{GOE} = 0.53$.

\section{Discussion and conclusions}
\label{sec:conclusions}

In these proceedings, we analyzed in detail the transition between the low-energy, low-dimensional eigenstates and the high-energy chaotic bulk of the spectrum for a simple supersymmetric quantum-mechanical model which mimics the structure of the BFSS matrix model \cite{Nicolai:NPB1989}. In our work \cite{Buividovich:22:2} we found that the low-energy eigenstates in this model support the growth of out-of-time-order correlators at low energies, even though they do not show any apparent signatures of quantum chaos. A proof of the continuity of the energy spectrum of the BFSS model, presented by de Wit, Lüscher and Nicolai in \cite{Nicolai:NPB1989} and illustrated using the simple model (\ref{HS}), suggests that similar low-energy eigenstates may also exist in the BFSS model. Given that at very low temperatures the BFSS model is expected to be holographically dual to a Schwarzschild black hole in $M$-theory, such low-energy eigenstates should presumably saturate the MSS bound on quantum Lyapunov exponent, making the system ``maximally chaotic''. It is therefore important to understand possible structure of low-energy eigenstates of the BFSS model. In these Proceedings, we presented a more detailed analysis of the properties of low-energy eigenstates which complements the results presented in \cite{Buividovich:22:2}.

Our first conclusion is that a sharp transition between the regular behavior of low-energy eigenstates and the chaotic, semi-classical behavior of higher-energy eigenstates does not lead to noticeable irregularities in the thermodynamic equation of state. It might well be that this transition only manifests itself in real-time quantities. Of course, thermodynamic singularities might still appear in the large-$N$ limit, for example, as higher-order phase transitions of Gross-Witten-Wadia type \cite{Gross:80:1}.

We also considered real-time spectral form-factors $K\lr{t} = \tr\lr{e^{i \hat{H} t - \hat{H}/T}}$ at high temperatures $T$ and found signatures of the universal ``ramp'' behavior \cite{Yoshida:1706.05400,Hanada:1803.08050} for both the supersymmetric and the bosonic Hamiltonians.

Our analysis of the entanglement entropy and entanglement spectrum of higher-energy eigenstates revealed some more nontrivial features of the low-energy to high-energy bulk transition. At high energies, the entanglement entropy $S_A\lr{E}$ behaves similarly in both the bosonic and the supersymmetric model, first growing with energy, and then saturating close to the Page entanglement entropy for a typical random (with respect to the Haar measure) state. This saturation supports the Eigenstate Thermalization Hypothesis (ETH) for the high-energy bulk of the spectrum. However, the entanglement entropy is practically energy-independent for the low-energy eigenstates of the supersymmetric model. The sharp transition between the low- and the high-temperature regimes seen in the $r$-ratio plot on Fig.~\ref{fig:r_ratios} is also seen as a rather sharp onset of the growth of entanglement entropy with energy. For a system with a holographic dual description, such as the BFSS model, the weak energy dependence of $S_A\lr{E}$ would imply that the geometry of the dual space-time does not change much upon adiabatic pumping of energy into the system. It would be interesting to understand what is the physical interpretation of this property, in particular, for the $M$-theory Schwarzschild black hole background that might be dual to low-energy part of the spectrum \cite{Maldacena:hep-th/9802042,Hanada:2110.01312}.

We also found some nontrivial features in the entanglement spectra of both the bosonic and the supersymmetric Hamiltonians. In both cases, the ground-state entanglement spectra are regularly spaced at low values of ``entanglement energies'' $\epsilon$, and start exhibiting universal random-matrix type level spacing fluctuations towards large $\epsilon$. This observation seems especially nontrivial for the supersymmetric model, as even the very regular low-energy states appear to still contain some information about the nearly-classical chaotic behavior of the system at high energies. Finally, we demonstrated that for eigenstates deep in the bulk of the spectrum of the original Hamiltonians $\hat{H}_B$ and $\hat{H}_S$ the entanglement spectrum is also well described by the universal Gaussian Orthogonal Ensemble of random matrices.

\acknowledgments{The author is grateful to Georg~Bergner, Masanori~Hanada, Piotr~Korcyl, Jacek~Wosiek and Jakub~Zakrzewski for interesting discussions, and to Masaki Tezuka for kindly providing numerical data on the energy spectrum of the SYK model. This work used the DiRAC@Durham facility managed by the Institute for Computational Cosmology on behalf of the STFC DiRAC HPC Facility (www.dirac.ac.uk). The equipment was funded by BEIS capital funding via STFC capital grants ST/P002293/1, ST/R002371/1 and ST/S002502/1, Durham University and STFC operations grant ST/R000832/1. DiRAC is part of the National e-Infrastructure. This work was performed using the DiRAC Data Intensive service at Leicester, operated by the University of Leicester IT Services, which forms part of the STFC DiRAC HPC Facility (www.dirac.ac.uk). The equipment was funded by BEIS capital funding via STFC capital grants ST/K000373/1 and ST/R002363/1 and STFC DiRAC Operations grant ST/R001014/1. DiRAC is part of the National e-Infrastructure.}


\end{document}